\documentclass[12pt]{article}

\usepackage{amsmath,epsfig,ulem}

\newcommand{\se}{{\tilde{e}}}

\newcommand{\st}{{\tilde{\tau}}}
\newcommand{\stR}{{\tilde{\tau}_1}}

\newcommand{\sn}{{\tilde{\nu}}}
\newcommand{\sne}{{\tilde{\nu}_e}}
\newcommand{\snl}{{\tilde{\nu}_l}}

\newcommand{\msn}[1]{m_{\tilde{\nu}_{#1}}}

\newcommand{\cha}{\tilde{\chi}}
\newcommand{\neu}{\tilde{\chi}^0}
\newcommand{\mcha}[1]{m_{\tilde{\chi}^\pm_{#1}}}

\newcommand{\hh}[1][1]{\tfrac{#1}{2}}

\newcommand{\PL}{\mathswitchr L}
\newcommand{\PR}{\mathswitchr R}

\def\mathswitch#1{\relax\ifmmode#1\else$#1$\fi}
\def\mathswitchr#1{\relax\ifmmode{\mathrm{#1}}\else$\mathrm{#1}$\fi}
\newcommand{\PW}{\mathswitchr W}
\newcommand{\PZ}{\mathswitchr Z}

\newcommand{\MW}{\mathswitch {M_\PW}}
\newcommand{\MZ}{\mathswitch {M_\PZ}}

\newcommand{\scrs}{{}}
\newcommand{\sw}{\mathswitch {s_{\scrs\PW}}}
\newcommand{\cw}{\mathswitch {c_{\scrs\PW}}}

\newcommand{\gev}{\,\, \mathrm{GeV}}

\newcommand{\SLASH}[2]{\makebox[#2ex][l]{$#1$}/}

\newcommand{\Eslash}{\SLASH{E}{.5}\,}
\newcommand{\RR}{{\rm R}}
\newcommand{\LL}{{\rm L}}

\newcommand{\anc}{\rule{0mm}{0mm}}

\newcommand{\mycaption}[1]{\caption{\sl #1}}

\begin{document}
\thispagestyle{empty}

\def\thefootnote{\fnsymbol{footnote}}

\begin{flushright}
DESY 05--123\\
FERMILAB--Pub--05/319--T\\
IFIC/05-26 \\
ZU-TH 09-05 \\
\end{flushright}

\vspace{1cm}

\begin{center}

{\Large\sc {\bf Determining Sneutrino Masses \\[2mm] 
                and Physical Implications}}        \\[3.5em]
{\large
\sc
A.~Freitas$^{1}$, W.~Porod$^{2,3}$ and P.~M.~Zerwas$^{4}$ 
}

\vspace*{1cm}

{\sl
$^1$ Fermi National Accelerator Laboratory, Batavia, IL 60510-500, USA

\vspace*{0.4cm}

$^2$ Universitat de Val\'encia, 46071 Val\'encia, Spain\\

\vspace*{0.4cm}

$^3$ Universit\"at Z\"urich, CH--8057 Z\"urich, Switzerland

\vspace*{0.4cm}

$^4$ Deutsches Elektronen-Synchrotron DESY, D--22603 Hamburg, Germany
}

\end{center}

\vspace*{2.5cm}

\begin{abstract}

In some areas of supersymmetry parameter space, sneutrinos are
lighter than the charg\-inos and the next-to-lightest neutralino, and they
decay into the invisible neutrino plus lightest-neutralino channel with
probability one. In such a scenario they can be searched for in decays
of charginos that are pair-produced in $e^+e^-$ collisions, and in associated
sneutrino-chargino production in photon-electron collisions. The sneutrino
properties can be determined with high accuracy from the edges of the decay
energy spectra in the first case and from threshold scans in the second.
In the final part of the report we investigate the mass difference of
sneutrinos and charged sleptons between the third 
and the first two generations in seesaw-type models
of the neutrino/sneutrino sector. For a wide range 
these mass differences are sensitive to the seesaw scale. 

\end{abstract}

\def\thefootnote{\arabic{footnote}}
\setcounter{page}{0}
\setcounter{footnote}{0}

\newpage


\section{Introduction}

The complex structure observed in the neutrino sector
will have interesting consequences for the properties of the sneutrinos,
the scalar supersymmetric partners of the neutrinos. These novel elements
require the extension of the minimal supersymmetric 
Standard Model MSSM, e.g.~by a superfield including the
right-handed neutrino field and its scalar partner \cite{Babu:1992ia}.  
In particular, if the small neutrino masses are generated 
by the seesaw mechanism \cite{seesaw}, 
a similar type of spectrum is induced in the scalar sector, splitting
into light TeV-scale and very heavy masses. Moreover, 
the intermediate seesaw scales will affect 
the evolution of the soft 
mass terms which break the supersymmetry 
at the high [GUT] scale,
particularly in the third generation
with large Yukawa couplings,  
so that universality will be broken at the low [electroweak] scale. 
This in turn will provide the opportunity to measure the
intermediate seesaw scale of the third generation indirectly under
well defined assumptions.                      
Two sets of observables are of central interest:  
\begin{itemize}
\setlength{\itemsep}{0ex}
\setlength{\parskip}{.1ex}
\item[--]
 The masses of the charged and neutral sleptons within the first two 
generations are determined by the soft scalar mass parameters and the
gaugino mass parameters at the 
unification scale, and the D-terms generated by the breaking of the 
grand unification and electroweak gauge symmetries.
\vspace{2mm}
\item[--]
 The mass differences between the sleptons 
of the third and the first two generations which are affected by the 
Yukawa couplings in the tau sector and the seesaw scale.
\end{itemize}
To develop a comprehensive picture, both the charged and the neutral sleptons
must be analyzed with high precision in parallel.\\

Sneutrinos can be pair-produced in $e^+e^-$ collisions. If they are
heavier than the light chargino and the next-to-lightest neutralino,
they can be searched for in the decays $\snl \to \nu_l \, \neu_2$ and
$\snl \to l^- \, \cha^+_1$, since these final states generate visible
signals in the detector. These channels have recently been studied in
Refs.~\cite{sneu,nauenberg} in detail. In contrast, if sneutrinos are
lighter than these particles, they decay only to final states $\snl
\to \nu_l \, \neu_1$ that are invisible and pair-production is useless
for studying these particles. [Photon tagging of these pairs remains
very difficult due to the reduced cross sections.]

However, in this configuration two other methods provide opportunities 
to study sneutrino masses:
\begin{itemize}
\setlength{\itemsep}{0ex}
\setlength{\parskip}{.1ex}
\item[--]
Chargino decays to sneutrinos and leptons,
\begin{equation}
\cha^\pm_1 \to l^\pm \, \sn_l^{(*)},
\label{eq:chadec}
\end{equation}
  with the charginos pair-produced in $e^+e^-$ annihilation. These two-particle
  decays develop sharp edges at the endpoints of the lepton energy spectra.
  Sneutrinos of all the three generations can be explored this way.
\vspace{2mm}
\item[--]
The first-generation sneutrino can also be studied in $e\gamma$ scattering
\cite{bhk}:
\begin{equation}
e^-\gamma \to \sne \cha^-_1. \label{eq:egprod}
\end{equation} 
The spectrum of Compton-backscattered laser light has a sharp peak at
  the maximum energy of the produced high-energy photons. In addition,
  the cross section
  for polarized $e,\gamma$ beams rises steeply at threshold, so that the scanning 
  of the threshold region can be used to determine the sum of the $\sne$ and
  $\cha^\pm_1$ masses. Since, on the other hand, the mass of
  $\cha^\pm_1$ can be measured in other channels with high precision, the
  process \eqref{eq:egprod} can serve as a channel for the $\sne$ 
  mass measurement. However, standard $W$ production gives rise to a serious
  background problem and the photon spectrum must be controlled very carefully. 
\end{itemize}
The slight shift of the parameters in the Snowmass point from SPS1a to SPS1a$'$
\cite{sps,spa}{\footnote{The shift lowers the prediction for the cold dark matter
density to the WMAP band.}} leads to a configuration in which the sneutrinos are 
lighter than the
charginos and the second-lightest neutralino. We therefore have adopted
the reference point SPS1a$'$ for carrying out detailed analyses of sneutrino
masses at a future linear collider \cite{lcreports}.

\vspace{1ex}
The material presented in this report is divided into three sections. 
In the next section we develop the
techniques for measuring the sneutrino masses in chargino decays, 
and in the subsequent section for
scanning the threshold region of the inelastic SUSY Compton process.
Detailed estimates of the expected errors on the sneutrino masses are
presented in the reference point SPS1a$'$ for both methods. In the final section
some interesting physics implications are worked out. First, the 
D-terms are estimated from the mass difference between the charged 
and neutral sleptons. Going beyond the MSSM in the next step, we discuss 
in particular the variation 
of the mass differences between sleptons of the third and the first two
generations with the Yukawa couplings and the intermediate seesaw scale.
This analysis 
is carried out for a minimal SO(10) grand unification scenario 
with universal boundary conditions 
for the soft scalar mass parameters, but potentially modified by D-terms  
associated with the GUT gauge symmetry breaking. In this configuration 
the system of charged and neutral slepton masses determines the seesaw
scale in addition to the complete set of universal soft mass parameters 
and D-terms at the GUT scale.


\section{Chargino Decays to Sneutrinos}
\label{dec}

\renewcommand{\arraystretch}{1.2}
\begin{table}[tb]
\begin{center}
\begin{tabular}{|c||c|c|r@{\:}ll|}
\hline
Sparticle & Mass $m$ [GeV] & Width $\Gamma$ [GeV] 
          & \multicolumn{3}{c|}{Decay modes} \\
\hline \hline
$\snl = \sne/\sn_\mu$ & $169.6$ & $0.09\,\,\,\,\,\,$ &
                $\snl$ & $\to \nu_l \, \neu_1$ & 100\%  \\
\hline
$\sn_\tau$ & $167.8$ & $0.15\,\,\,\,\,\,$ &
                $\sn_\tau$ & $\to \nu_\tau \, \neu_1$ & 100\% \\
\hline
$\st_1$ & $105.7$ &  $0.0037$ & $\st_1$ & $\to \tau \, \neu_1$ & 100\% \\
\hline \hline
$\neu_1$ & $100.8$ & --- & \multicolumn{2}{c}{---} & \\
\hline
$\cha_1^\pm$ & $180.5$  & $0.074$ &
        $\cha^+_1$ & $\to \stR^+ \, \nu_\tau$ & 53\% \\
        &                     & && $\to \sn_e \, e^+$ & 13\% \\
        &                     & && $\to \sn_\mu \, \mu^+$ & 13\% \\
        &                     & && $\to \sn_\tau \, \tau^+$ & 19\% \\
\hline
\end{tabular}
\end{center}
\vspace{-1em}
\mycaption{Tree-level masses, widths and main branching ratios of sleptons and of the 
light neutralino and chargino states at Born level
for the reference point SPS1a$'$ \cite{spa}.}
\label{tab:sps1}
\end{table}
The relevant masses, widths and branching ratios for the reference point
SPS1a$'$ are  listed in Tab.~\ref{tab:sps1}.
In this reference point the branching ratio for the 2-body decay
$\cha^\pm_1 \to \sn_l^{(*)} \, l^\pm$ $(l = e,\mu)$
of the lightest chargino amounts to 13.4\%. With a production cross
section of 100--200 fb not far above the threshold, a large number of
sneutrinos can be generated.  For a given $e^+e^-$
energy the lepton decay energies are almost uniformly distributed
between the minimum and maximum value $E_{\rm min}$ and $E_{\rm max}$:
\begin{equation}
E_{\rm min,max} = \frac{\sqrt{s}}{4} \; \frac{\mcha{1}^2-\msn{l}^2}{\mcha{1}^2}
\; \left( 1 \pm \sqrt{1- 4 \mcha{1}^2/s} \right). \label{eq:edges} \\
\end{equation}
These values determine the masses of the
parent chargino and the child sneutrino:
\begin{eqnarray}
\mcha{1} &=& \sqrt{s} \,\frac{\sqrt{E_{\rm min} E_{\rm max}}}{E_{\rm min} + 
	E_{\rm max}},
\label{eq:mcha} \\
\msn{l} &=& \mcha{1} \sqrt{1- \frac{2(E_{\rm min} + E_{\rm max})}{\sqrt{s}}}\,.
\label{eq:msnu}
\end{eqnarray}
At the threshold for the production of the chargino pair the
energy distribution is reduced to a sharp line.

The uniform distribution is distorted only slightly by polarization effects.
Close to threshold the produced chargino $\cha^\pm_1$ is longitudinally and
transversely polarized in the production plane with degrees \cite{choi:00}
\begin{equation}
     P_{\rm L} = - P_0  \cos\theta \;\;\;{\rm{and}}\;\;\;
                          P_{\rm \perp} = + P_0 \sin\theta
\end{equation}
respectively, where $\theta$ denotes the polar production angle.
The polarization degree $P_0$,
\begin{equation}
       	  P_0 = \frac{Q_\PL^2 - Q_\PR^2}{Q_\PL^2 + Q_\PR^2}, \\
\end{equation}
can be expressed by the bilinear L/R charges
\begin{align}
  Q_\PL &= 2 + \frac{D_\PZ}{\sw^2\cw^2} (\sw^2 -\hh) (2\sw^2 - |U_{11}|^2 -
   \hh |U_{12}|^2 - |V_{11}|^2 - \hh |V_{12}|^2)
   - \frac{D_\sn}{4\sw^2}, \\
  Q_\PR &= 2 + \frac{D_\PZ}{\cw^2} (2\sw^2 - |U_{11}|^2 -
   \hh |U_{12}|^2 - |V_{11}|^2 - \hh |V_{12}|^2).
\end{align}
Here, $D_\PZ = s / [s - \MZ^2 + i \MZ \Gamma_\PZ]$ and  $D_\sn = s /
[t-\msn{}^2]$ are the $Z$-boson s-channel and the ${\tilde{\nu}}_e$ t-channel
exchange propagators, respectively, renormalized by the energy squared $s$. 
$U_{ij}$ and $V_{ij}$ are the mixing matrices of the negatively/positively
charged charginos; for details see Ref.~\cite{slep}.

With respect to the polarization axis the angular distribution of the lepton in
the decay \eqref{eq:chadec} of a completely polarized chargino 
must follow the $\cos\theta^\ast$ law 
as a consequence of angular momentum conservation, 
\begin{equation}
\frac{4\pi}{\Gamma} \; \frac{{\rm d}\Gamma}{{\rm d}\cos\theta^\ast} = 
1 - \cos\theta^\ast,
\end{equation}
where 
$\theta^\ast$ is the lepton polar angle with respect to the chargino
polarization axis in the chargino rest frame.
Right-handedly polarized charginos decay to left-handedly polarized leptons for
spin-flipping scalar couplings so that backward emission is dominant. 

Near the threshold the degree $P_0$ of the chargino is close to unity. However,
the longitudinal polarization which affects the lepton energy distribution, 
averages approximately to zero if integrated over the chargino production
angle. Farther away from the threshold small polarization 
effects build up slowly. They can be calculated quantitatively
by adopting the general analysis from Refs.~\cite{Zstop,choi:00}.

\vspace{1ex}
{\bf (A)} Assuming the two first and second generation L-sneutrinos,
 ${\tilde{\nu}}_e$ and ${\tilde{\nu}}_{\mu}$, to be
mass-degenerate, a final state $e \, \mu + \Eslash$, including one electron,
one muon and missing energy carried away by the neutrinos and neutralinos,
is little contaminated
by backgrounds. The primary background source from Standard Model processes 
are $W^+W^-$ pair production and
single $W$ production, with the
$W$'s decaying into a charged lepton and the associated neutrino,
$W^\mp \to l^\mp \! \stackrel{_{_{(-)}}}{\nu_l}$.

\renewcommand{\arraystretch}{1.5}%
\begin{table}[tb]
\centering
\begin{tabular}{|p{6.2cm}|p{5.3cm}|p{3.7cm}|}
\hline
Condition & Variable & Accepted range \\
\hline \hline
Reject leptons in forward/\newline backward region
& lepton polar angle $\theta_{
\rm l}$ &
  $|\cos \theta_{\rm e}| < 0.90$ \newline
  $|\cos \theta_{\rm \mu}| < 0.95$ \\
Reject soft leptons from radiative photon splitting and
$\gamma$-$\gamma$ events
 & lepton energy $E_{\rm l}$ & $E_{\rm l} > 5$ GeV
\\
Reject missing momentum in forward/backward region
from particles lost in the beam pipe &
  missing momentum polar \newline angle $\theta_{\vec{p}_{\rm miss}}$ &
  $|\cos \theta_{\vec{p}_{\rm miss}}| < 0.90$ \\
Angular separation between \newline $e$ and $\mu$ lepton &
  angle $\phi_{\rm e\mu}$ between electron \newline and muon
  & $|1- \cos \phi_{\rm e\mu}| > 0.015$ \\
\hline
\end{tabular}
\mycaption{Cuts to reduce the main Standard Model backgrounds and
to account for the detector geometry and resolution.}
\label{tab:cuts}
\end{table}
\renewcommand{\arraystretch}{1}%
Due to spin correlations and the boost factor, the background from $W$
bosons tends to peak along the beam direction and can be reduced by
requiring both charged leptons in the central detector region. A
simple set of cuts, summarized in Tab.~\ref{tab:cuts}, also reduces
backgrounds from soft and collinear photon contributions and tau-induced 
backgrounds below the per-cent level.

Besides direct chargino and $W$-boson decays into electrons and
muons, the $e \mu + \Eslash$ signature can also be generated by leptonic tau
decays. In the numerical analysis, cascade decays via intermediate tau leptons
have been included both for the chargino production process and the $W$
background.

Several other supersymmetric channels are open which give rise to $e
\mu + \Eslash$ final states through tau decays.  The cross section for
$\tilde{\tau}$ pair production, with $\tilde{\tau}$ decaying to $\tau
{\tilde{\chi}}^0_1$, is of similar size as the signal cross section
after the leptonic tau branchings are included; however, cascading
down to the $e$ and $\mu$ final leptons in two steps, the energies are
softened to small values. Therefore we do not expect this channel to
have a significant impact on the determination of neither the upper
nor the lower edges of the signal lepton distributions. Nevertheless,
this channel has been included explicitly in the numerical
calculation. Cascades from chargino and neutralino pairs to $e,\mu$
leptons through taus are doubly suppressed by the leptonic tau
branching ratios, cf.~Ref.\cite{CMZ}, and they involve at least two
consecutive decay steps so that the energy distribution is doubly
soft. These background channels could therefore be neglected in the
present analysis.\\

\begin{figure}[p]
\raggedright \anc \hspace{13mm} (a) \hspace{2.7in} (b)\\[1ex]
\centering
\epsfig{file=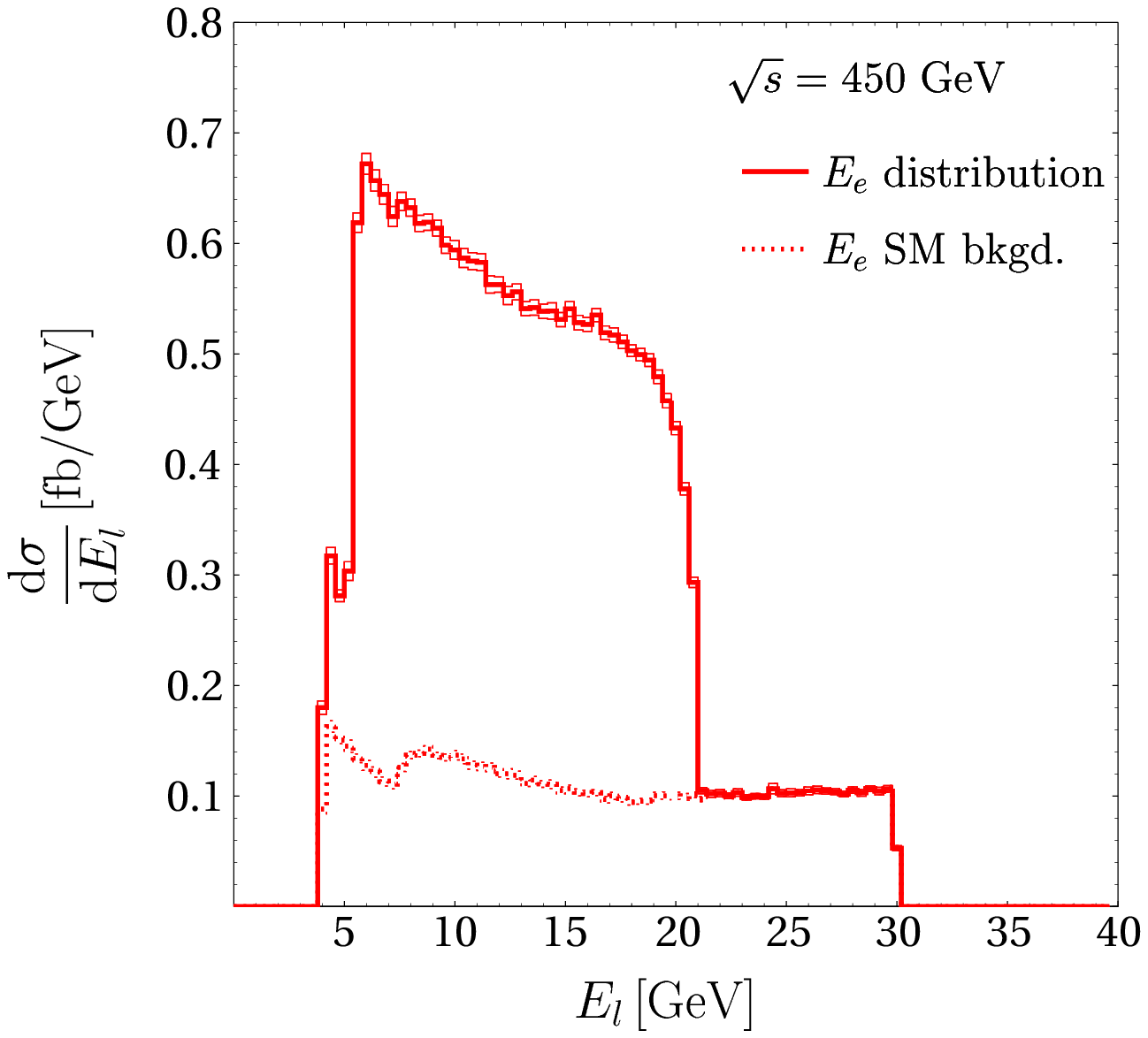,width=3.3in, bb=8 345 375 681, clip=true}
\epsfig{file=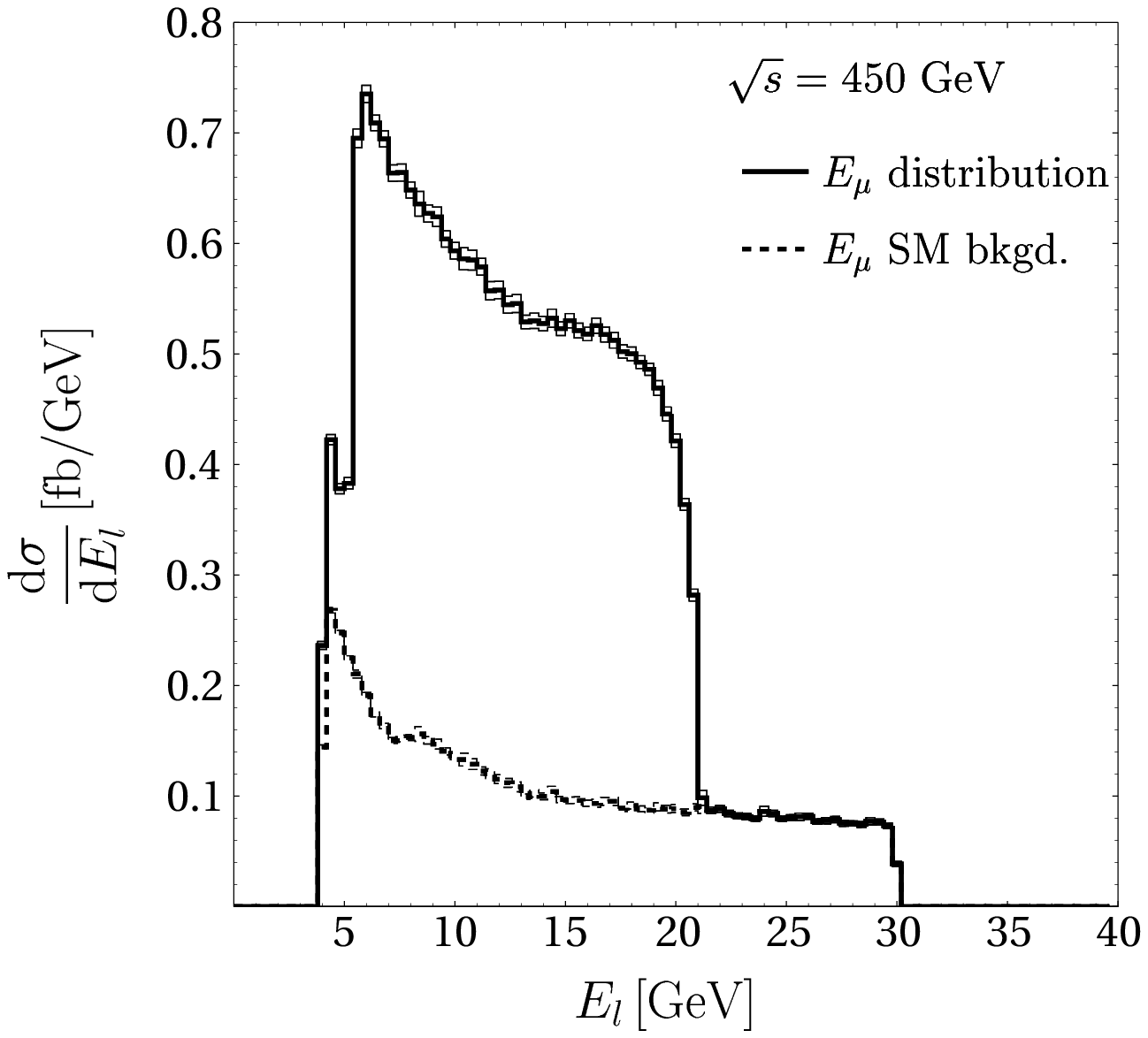,width=2.92in, bb=50 345 375 681, clip=true}
\mycaption{(a) Electron and (b) muon energy distributions 
for the process $e^+e^- \to (\cha^+_1
\cha^-_1) \to e^\pm \mu^\mp \sn_e \sn_\mu \to e^\pm \mu^\mp +
E\hspace{-1.3ex}/$. The signals for the lepton energy spectra are shown 
over the background from Standard Model
sources. Cuts of 4 GeV and 30 GeV are included for the lepton
energies.}
\label{fig:specl}
\end{figure}

Results for the lepton energy distributions are shown in Fig.~\ref{fig:specl}
for the signals and the backgrounds. The energy has been chosen at 450 GeV, where
the cross section is maximal, and an integrated luminosity of 500 fb$^{-1}$ 
has been assumed.  
\begin{figure}[p]
\raggedright \anc \hspace{13mm} (a) \hspace{2.7in} (b)\\[1ex]
\centering
\epsfig{file=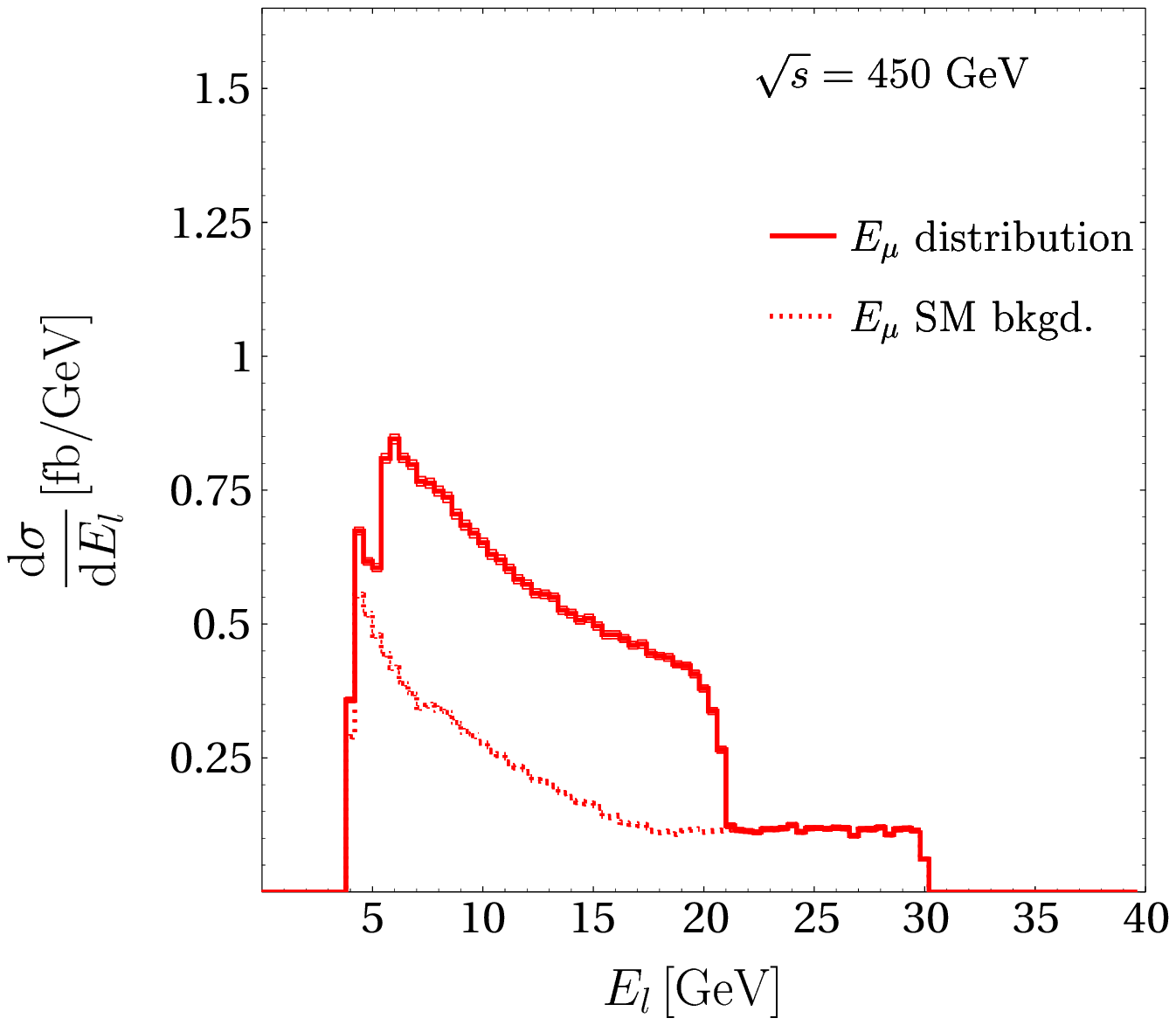,width=3.37in, bb=8 345 390 681, clip=true}
\epsfig{file=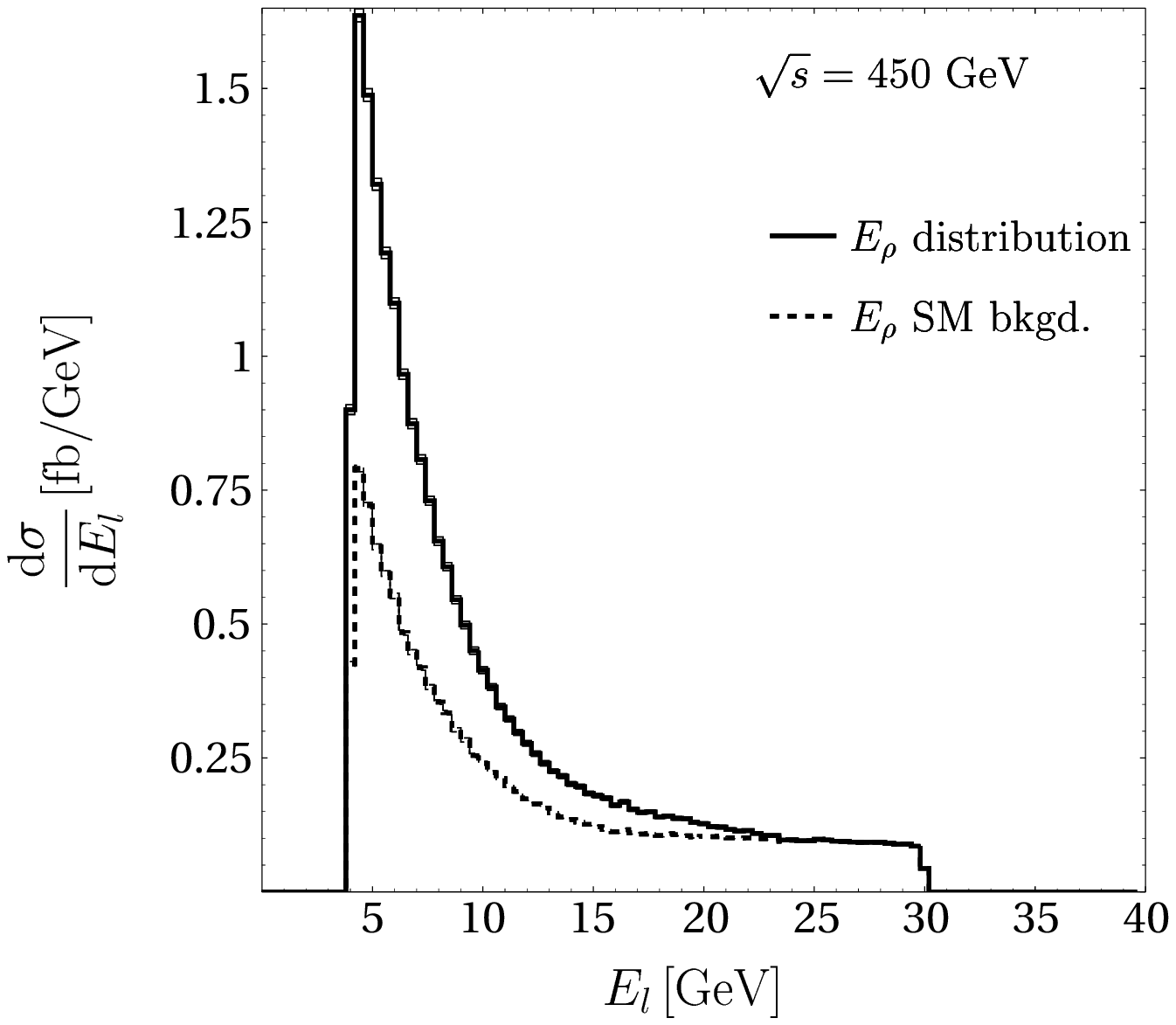,width=3.0in, bb=50 345 390 681, clip=true}
\mycaption{(a) Muon and (b) tau hadron energy distributions for the process
$e^+e^- \to (\cha^+_1
\cha^-_1) \to \mu^\pm \tau^\mp \sn_\mu \sn_\tau \to \mu^\pm \tau^\mp +
E\hspace{-1.3ex}/$. As in Fig.~\ref{fig:specl}, Standard Model backgrounds are
shown seperately and 
cuts of 4 GeV and 30 GeV are included for the lepton
energies.}
\label{fig:spect}
\end{figure}
The signal spectrum shows the expected almost uniformly flat distribution. The
edges are slightly rounded due to initial-state radiation and beamstrahlung.
Backgrounds from $W$-boson production are relatively small and flat so that
the kinematical edges of the signal distribution are not distorted. The
position of the edges $E_{\rm min}$ and $E_{\rm max}$ can be extracted 
from a simple fit using a box-shaped fit function convoluted with the
initial-state radiation and beamstrahlung spectra.
For the reference point SPS1a$'$ the fit to the electron energy spectrum yields
well determined values from which the chargino and sneutrino masses can be derived:  
\begin{align}
  E_{\rm min,e} &= 5.30^{+0.07}_{-0.06} \mbox{ GeV}, &
  E_{\rm max,e} &= 21.04^{+0.07}_{-0.07} \mbox{ GeV},               \\
\mcha{1} &= 180.4^{+0.7}_{-0.7} \mbox{ GeV}, &
\msn{e} &= 169.5^{+0.7}_{-0.6} \mbox{ GeV}.
\end{align}
Similarly the muon energy spectrum is fitted with the results
\begin{align}
  E_{\rm min,\mu} &= 5.30^{+0.08}_{-0.07} \mbox{ GeV}, &
  E_{\rm max,\mu} &= 21.06^{+0.07}_{-0.07} \mbox{ GeV},              \\
\mcha{1} &= 180.4^{+0.8}_{-0.7} \mbox{ GeV}, &
\msn{\mu} &= 169.5^{+0.8}_{-0.7} \mbox{ GeV}.
\end{align}

Note that the accuracy of the chargino mass corresponds to the value
expected from a threshold scan \cite{lhclc}, $\delta_{\rm thr} \mcha{1} = 
0.55$ GeV.
Combining the two methods of chargino mass measurement, 
the accuracies sharpen to
\begin{equation}
\delta \mcha{1} \approx \; \pm 0.45 \mbox{ GeV} \qquad \mbox{and} 
\qquad
\delta \msn{e,\mu} \approx \; \pm 0.40 \mbox{ GeV}.
\end{equation}
Thus a very high accuracy can be reached in this configuration.

\vspace{1ex}
{\bf (B)}
A special case are chargino decays to $\sn_\tau$ and $\tau$. Since the tau
decays into an invisible neutrino and a charged pion, rho meson or other
low-mass hadrons, the energy distribution of the visible meson is
not uniform but follows asymptotically the familiar relation $\log(1/x_{\rm had})$ 
for the (observed) fraction $x_{\rm had}$ of the $\cha^\pm_1$ energy. 
This distribution
does not generate a sharp edge anymore as in uniform energy distributions
so that the error in $\sn_\tau$ mass measurements increases
correspondingly.                                                                

In the following, all hadronic tau decay channels with a total
branching ratio of 65\% are included. For simplicity, the tau decay
kinematics is approximated by the dominant decay into a rho meson.
The upper tail of the decay energy spectrum for other important
hadronic tau decay modes into one or three pions or into the $a_0$
meson are very similar to the rho meson spectrum, so that this
approximation is sufficient for the purpose of this analysis. The same
standard cuts listed in Tab.~\ref{tab:cuts} for the muon are applied
for the final state rho meson, only with a stricter cut on the polar
angle, $|\cos
\theta_{\rm \rho}| < 0.8$, $E_{\rm \rho} > 5$ GeV, $|1- \cos \phi_{\rm l\rho}|
> 0.015$. The upper endpoint of the rho meson energy spectrum is the same as
the maximum energy $E_{\rm max,\tau}$ of the tau coming from the chargino
decay, as given by eq.~\eqref{eq:edges}. 

In addition to the decay of the charginos into $\sn_\tau$ and $\tau$, the
decay into a charged stau and neutrino, $\cha^\pm_1 \to \st^\pm_1 \,
\stackrel{_{_{(-)}}}{\nu_\tau}$, is also possible in the SPS1a$'$ scenario (see
Tab.~\ref{tab:sps1}). However, the maximum energy of a tau lepton originating
from the latter decay is smaller,
about 15.8 GeV. This value is well below the upper endpoint of the tau energy spectrum 
from the sneutrino decay chain, $\cha^\pm_1 \to \sn_\tau^{(*)} \, \tau^\pm$, $E_{\rm
max,\tau} = 24.5$ GeV, so that this endpoint is not contaminated by the $\st_1$
background{\footnote{This is generically expected if the $\st_1$--$\neu_1$ mass
difference is significantly
smaller than the $\sn_\tau$--$\cha^\pm_1$ mass difference.
In other scenarios, however, the chargino decays into staus might
make the extraction of the tau-sneutrino signal virtually impossible.}}. 

For analyzing the tau-sneutrino decay chain, it is expedient to consider 
the final state $\mu \, \tau + \Eslash$, which has lower background levels than the
$e \, \tau + \Eslash$ final state, since the single $W$ process does not
contribute.

As before, only ${\tilde{\tau}}$ pair production is taken into account
explicitly as a supersymmetric background channel. Neutralino pairs
leading to $\tau \mu + \Eslash$
are suppressed and accumulate at small energies. However, as the tau
signal channel does not generate sharp edges in the visible $\rho$
energy distribution anymore, we focus the analysis on the clean upper
onset of the $\rho$ spectrum. Since all the parameters of the
background $\tilde{\tau}$ channel can be pre-determined independently with
high precision \cite{martyn}, this contribution can be calculated
reliably.

The energy distributions for the muon and the
hadronic tau decay products [assuming the decay $\tau \to \rho \nu_\tau$] are
shown in Fig.~\ref{fig:spect} for signal and Standard Model 
and supersymmetric background channels.\\

A fit to the upper edge of muon energy spectrum yields 
$E_{\rm max,\mu} = 21.0 \pm 0.1 \mbox{ GeV}$, in agreement with the previous
estimate.  The rho meson energy endpoint can be extracted by a fit
to Monte-Carlo template samples. From the upper tail of the rho meson
spectrum
one obtains for the energy spectrum endpoint
\begin{equation}
E_{\rm max,\tau} = 24.5^{+1.0}_{-1.0} \mbox{ GeV}.
\end{equation}
Using the chargino mass extracted from the muon distribution,
eq.~\eqref{eq:mcha}, as an input, the $\sn_\tau$ mass can be derived
from this endpoint,
\begin{equation}
\msn{\tau} = 167.6^{+0.9}_{-0.8} \mbox{ GeV}. 
\end{equation}
All errors quoted above are statistical errors only. Nevertheless, the 
results derived in this analysis are clearly encouraging.
[A detailed experimental simulation
including systematic errors is beyond the scope of our analysis.]


\section{\boldmath $e\gamma$ Production of Sneutrinos}

Electron sneutrinos can also be studied in associated production with
charginos in electron-photon collisions \cite{bhk}. The photon beam is generated through
laser light scattering on the second incoming electron beam \cite{ginz}.
If circularly polarized laser photons are back-scattered off electrons
of opposite-sign helicity, the spectrum of the generated high-energy photons 
has a sharp maximum at the kinematical edge,
\begin{equation}
E_{\gamma,\rm max} = \frac{E_{\rm b}}{1 + x^{-1}}
\end{equation}
with
\begin{equation}
x = \frac{4 E_0 E_{\rm b}}{m^2_e} \simeq 4.8 \,,
\end{equation}
where $E_0$ denoting the laser photon energy, $E_{\rm b}$ the electron beam
energy and $E_\gamma$ the resulting scattered photon energy; 
the ratio $E_{\gamma,\rm max} / E_{\rm b}$ is close to 0.8
for standard electron and laser energies. 
The high-energy
photons are circularly polarized themselves at the edge with helicity opposite
to the helicity of the laser photons.\\

If in the subsequent $e\gamma$ collision process  
electron and photon helicities are chosen of equal sign, the final-state system
is generated in an S-wave near the threshold. The cross section therefore rises 
sharply at the threshold proportional to the velocity $\beta$ of 
the sneutrino \cite{bhk}, 
\begin{equation}
\sigma [e^-_\LL \gamma_\LL \to \sne \cha^-_1]
 = \frac{2 \pi \alpha {\mcha{1}^2} \beta }{s^2} \, \frac{V^2_{11}}{\sw^2} \,
 \times \sum_{\lambda = \pm 1} (1 + \lambda\beta)
  \left[\sqrt{(1-\beta^2)s/\mcha{1}^2} - (1+\lambda\beta) \right]^2
 + \mathcal{O}(\beta^2).
\end{equation}
Thus traversing the energy threshold for the supersymmetric Compton process
leads to the onset of a striking increase of the observed production rate. If
the chargino mass has been pre-determined, the threshold energy,
\begin{equation}
  E_{\rm thr} = \mcha{1} + m_{\sne},
\end{equation}
can be exploited  to extract the electron-sneutrino mass.\\

This theoretical picture is modified in a more realistic analysis of the
spectrum of the back-scattered laser photons. During the beam collisions 
the electrons are strongly deflected by the opposite electron beam. 
As a result, the $e\gamma$ luminosity is reduced compared to the naive 
expectation without repulsion. Due to the deflection away from the axis, 
the electrons do not collide in general with the highest energy photons, 
but with those of somewhat lower energy. Furthermore, the simultaneous 
absorption of more than one laser photon generates additional peaks
at high photon energies.

These effects can be calculated by Monte Carlo beam simulations \cite{telnov}. 
We have adopted for our analysis the convenient parametrization of
Ref.~\cite{compaz}, 
which has been adjusted to detailed simulations and which
accounts well for the characteristics of the spectrum. 
The maximum energy of the primary peak is reduced compared with the 
value in the single Compton process, and a second but
small peak develops at high energies.
The primary edge is by far dominant and it is very pronounced so 
that on the whole a sharp onset is still guaranteed despite of the complicated
interactions between the electron and photon field. Moreover, the edge can be
calibrated by scanning the threshold for $e^-\gamma \to \se^-\neu_1$ 
production which can be predicted very accurately from $e^+e^-$ data and 
which can therefore be exploited to control 
the beam spread parameters near the maximum of the distribution.\\

The signal for associated sneutrino-chargino production in $e\gamma$ collisions
is characterized by a single isolated lepton coming from the chargino decay
\eqref{eq:chadec} plus missing energy. The muon final state,
\begin{equation}
e^-\gamma \to \sne \cha^-_1 \to \sne \, \sn_\mu^* \mu^-,
\end{equation} 
is less contaminated by backgrounds than the electron final state and leads
to experimentally cleaner final states than tau which involves subsequent 
decay channels. 
Moreover, the supersymmetric background channel $\tilde{e} {\tilde{\chi}}^0_1
\to e + {\tilde{\chi}}^0_1 {\tilde{\chi}}^0_1$
is eliminated from the $\mu$ sample. 
Therefore in the following only the muon final state will be
studied. Large backgrounds arise from single $W$-boson production,
\begin{equation}
e^-\gamma \to \nu_e W^- \to \nu_e \bar{\nu}_\mu \mu^-.
\end{equation} 
For the sneutrino signal, the muon energy is relatively small due to the small
mass difference $\mcha{1} - \msn{\mu}$, whereas the $W$ background leads to larger
values for the muon energy $E_\mu$. This distinction is somewhat washed out by
the photon energy spectrum, but a cut $E_\mu < 25$ GeV is still effective to
reduce the background drastically.
In addition, the decay muon from the $W$ boson is peaked in the forward
direction, $\cos\theta > 0$, where $\theta$ is the angle between the
direction of the incoming electron beam and the outgoing muon. Consequently, by applying
a cut on the scattering angle, $\cos\theta < 0$, the signal-to-background is
further improved. As before, it is also required that the
muon has a minimal energy of 5 GeV and is emitted into the central region of the
detector, $|\cos \theta_\mu| < 0.95$.\\

\begin{figure}[tb]
\centering
\epsfig{file=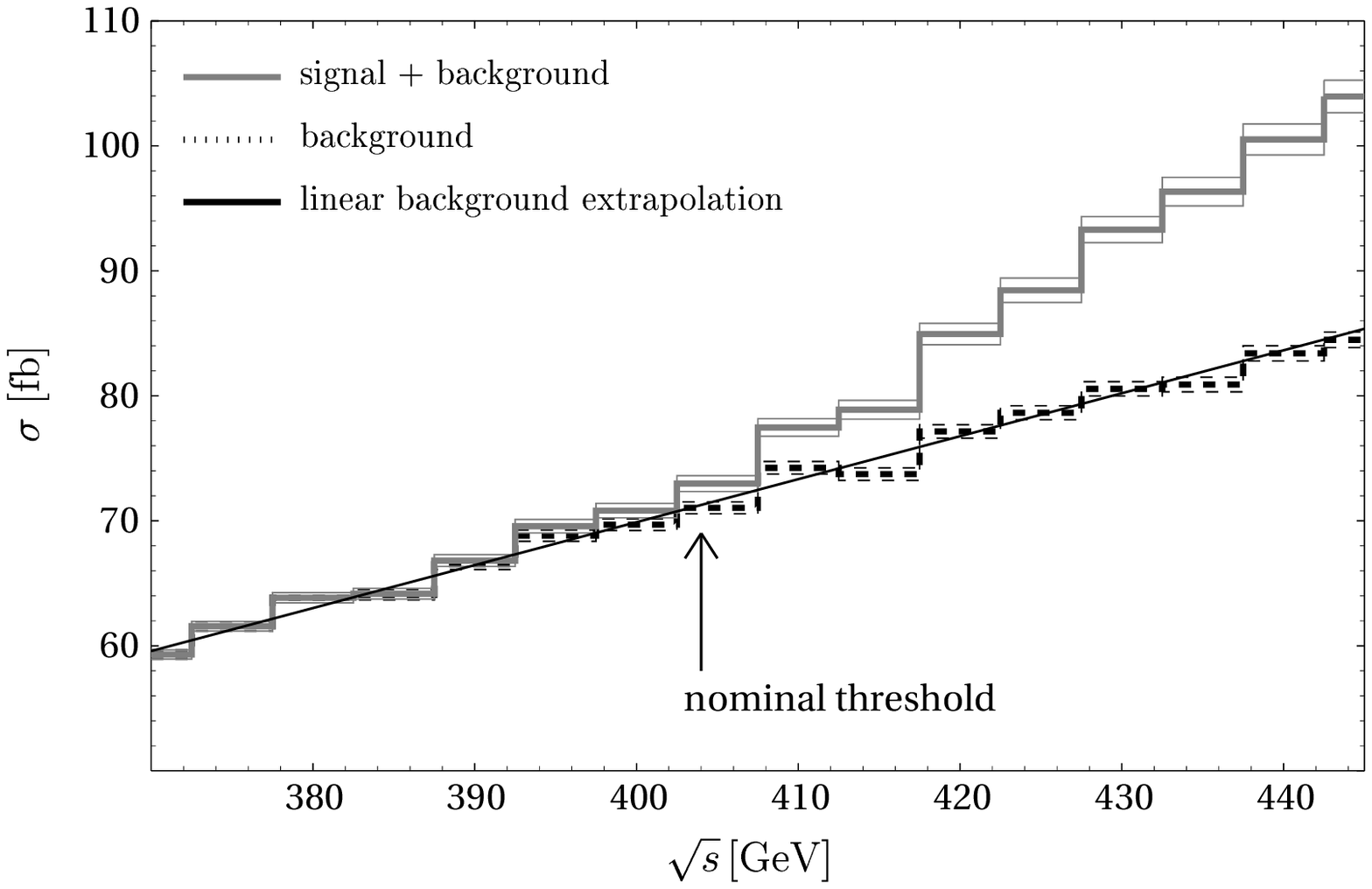, width=6in, bb=0 390 488 681}
\mycaption{The excitation curve for $e^-\gamma \to \cha^-_1 \sne$ production
over the dominant Standard Model background as a function of the center-of-mass
energy of the two incoming electron beams (one of which is used for generating
the high-energy photons). The plot has been generated
including the photon beam spectrum, beamstrahlung and initial-state-radiation
for the electron beam, polarization of both beams and cuts to reduce the
background. 
}
\label{fig:egthr}
\end{figure}

In Fig.~\ref{fig:egthr}, the onset of the sneutrino-chargino threshold
is shown after convolution with the photon energy spectrum of
Ref.~\cite{compaz}; beamstrahlung and initial-state radiation for the
$e^-$ beam are included. Also shown is the remaining background after
applying the cuts introduced above. As explained before, the signal is
enhanced by choosing left-polarized $e^-$ beams and right-handed
polarization for the laser photons, resulting in a left-polarized
photon beam near the kinematical edge. A polarization degree of 90\%
is assumed for the $e^-$ beams, while the laser source is taken to be
100\% polarized.

As evident from the figure, the background is still significantly larger than the
sneutrino-chargino signal, but the expected rates are large, thus allowing a
statistical distinction even in this situation. The dependence of the background
on the $e^-e^-$ center-of-mass energy $\sqrt{s}$ can be well described by a
linear approximation, so that it can be extrapolated from measurements at values
of $\sqrt{s}$ below the onset of the signal, as indicated in the figure.

We estimate the precision for the sneutrino mass measurement from a threshold
scan based on simulated data at six equidistant scan points in the range
$\sqrt{s} = 380 \; {\rm{to}} \; 430$ GeV for the initial $e^-e^-$ energy. 
An integrated luminosity of 10 fb$^{-1}$ is
assumed to be spent on each scan point. The sneutrino mass is extracted 
using a binned
likelihood method.  Taking into account the
chargino mass measurement error from a $\cha^+_1\cha^-_1$ threshold scan, 
$\delta_{\rm thr} \mcha{1} = 
0.55$ GeV, this leads to
\begin{equation}
\msn{e} = 169.8 \pm 3.2 \mbox{ GeV}. 
\end{equation}
While the precision achieved by this method is worse than the mass
determination from chargino decay spectra elaborated in section~\ref{dec}, the
$e\gamma$ threshold scan nevertheless provides an interesting alternative to
search for invisibly decaying  particles and to determine their
mass with per-cent level accuracy. It may also be viewed
as a complementary check of decay spectra measurements.

\section{Implications: Determining GUT parameters and the seesaw scale}

Complementing the precision measurements of the charged slepton masses
by the sneutrino masses of all three generations has exciting
consequences for exploring physics scenarios at high scales. This 
includes potentially the universality of the soft scalar mass parameters,
D-terms associated with the breaking of the grand unification gauge
theory, and last-not-least the size of the seesaw scale in neutrino 
physics.

We will illustrate these points in an SO(10) model{\footnote{The pattern of 
mass ratios among the supersymmetric particles remains analogous to the
SPS1a$'$ model so that measurement errors can be obtained by scaling from
the preceding sections.}} in which the matter superfields of the three 
generations belong to 
16-dimensional representations of SO(10) and the standard Higgs superfields 
to 10-dimensional representations while a Higgs superfield in the 
126-dimensional representation generates Majorana masses to 
the right-handed neutrinos. As a result, the Yukawa couplings 
in the neutrino sector
are proportional to the up-type quark mass matrix, for which
the standard texture is assumed. The SO(10) symmetry is  
broken to the Standard Model SU(3)$\times$SU(2)$\times$U(1) 
symmetry at the grand unification scale $M_U$ directly. We will assume 
universality 
for the sfermion mass parameters at $M_U$, yet potentially modified 
by D-terms 
of the order of the electroweak scale \cite{Kolda:1995iw}:
\begin{align}
   m^2_{\rm E} &= m^2_{16} + D_U,   \\ 
   m^2_\LL &= m^2_{16} - 3D_U,      \\ 
   m^2_\RR &= m^2_{16} + 5D_U.
\end{align}
$m_{\rm E}$, $m_\LL$ and $m_\RR$ denote the iso-singlet, -doublet and 
R-sneutrino scalar mass parameters, $m_{16}$ the universal scalar mass, 
and $D_U$ the SO(10) D-term contribution. 
For simplicity we identify the soft breaking masses
of the Higgs sector also with $m_{16}$; this technical simplification can 
easily be relaxed.

Assuming that the Yukawa couplings are the same for  up-type quarks
and neutrinos at the GUT scale and that the Majorana mass matrix of
the right-handed neutrinos has a similar structure, one obtains 
a (weakly) hierarchical neutrino mass spectrum and nearly bi-maximal mixing 
for the left-handed neutrinos  \cite{Buchmuller:2001dc}.\footnote{The neutrino
Yukawa couplings induce off-diagonal elements in $M^2_L$ and $A_l$ leading
to flavour changing sneutrino decays. However, as we will see later, these
are CKM suppressed and can, thus, safely be neglected in the following.}
In this class of models the masses of the right-handed neutrinos 
are also hierarchical{\footnote{These 
relations
are supposed to hold only at the level of very rough estimates;
the only crucial
point is the outstanding size of the $\RR_3$ mass.}},
\begin{equation}
   M_{\RR_3} :  M_{\RR_2} : M_{\RR_1} \sim m^2_t : (\kappa_c m_c)^2 : (\kappa_u m_u)^2
\label{eq:MRratios}
\end{equation}
with $\kappa$'s of order 10.
The overall scale is set by $m_t$ and 
the mass of the heaviest neutrino:
\begin{equation}
   M_{\RR_3} \sim m^2_t / m_{\nu_3}
\label{eq:MR3}
\end{equation} 
For $m_{\nu_3} \sim 5 \times 10^{-2}$ eV, the heavy neutrino mass of the
third generation amounts to $\sim 6 \times 10^{14}$ GeV, {\it i.e.} a value
close to the grand unification scale $M_U$. 

Even though the soft mass parameters are assumed to be of 
the order of the electroweak scale or slightly above, the seesaw 
mechanism is induced also in the scalar sneutrino sector through the
Higgs-126 couplings in the superpotential, see 
e.g.~\cite{Masiero:2002jn,blair:00p}, 
\begin{equation}
   m^2_{\tilde{\nu}_{k\RR}} \simeq m^2_{\RR_k}  
\end{equation}
so that R-sneutrinos cannot be generated at colliders.\\

The SO(10) gauge invariance requires a unified gaugino mass parameter $M_{1/2}$
at the scale $M_U$.
We choose the high-scale supersymmetry breaking parameters in the framework of
the SPS1a$'$ scenario,
\begin{equation}
M_{1/2} = 250 \gev, \quad
m_{16} = 70 \gev, \quad
A_0 = -300 \gev, \quad
D_U = 0
\end{equation}
and
\begin{equation}
\tan\beta = 10, \quad
\mu > 0.
\end{equation}

This minimalistic model is compatible with all the observations 
in the neutrino sector. It may serve to illustrate the potential
of precision measurements of charged and neutral scalar lepton masses 
for exploring physics close to grand unification/Planck scale.\\

To leading order of the solutions of the renormalization group 
equations{\footnote{One-loop
corrections are properly accounted for in the numerical analyses; 
they cancel to a large extent if mass differences are calculated.} 
the masses of the scalar selectrons and $e$-sneutrino can be expressed by 
the high scale parameters $m_{16}$ and $M_{1/2}$, and the D-terms:
\begin{align}
\label{eq:mselr}
 m^2_{\tilde{e}_R} &= m^2_{16} + D_U + \alpha_R M^2_{1/2}  
                                                   - s^2_W M^2_Z \cos 2 \beta, \\
 m^2_{\tilde{e}_L} &= m^2_{16} - 3 D_U + \alpha_L M^2_{1/2}
                              - (1/2 - s^2_W) M^2_Z \cos 2 \beta, \\
 m^2_{\tilde{\nu}_{eL}} &= m^2_{16}  - 3 D_U + \alpha_L M^2_{1/2}  
                                               + 1/2 M^2_Z \cos 2 \beta.
\end{align}
The coefficients $\alpha_R$ and $\alpha_L$ can be determined numerically
and the integration of the 1-loop RGEs yields  
$\alpha_R \simeq 0.15$ and $\alpha_L\simeq 0.5$.
Analogous representations can be derived, to leading order, for the scalar 
masses of 
the third generation, complemented however by additional 
contributions $\Delta_\tau$
and $\Delta_{\nu_\tau}[M_\RR]$ from the standard tau Yukawa term and the Yukawa 
term in the tau neutrino sector, respectively:
\begin{align}
\label{eq:mtaur}
 m^2_{\tilde{\tau}_R} &= m^2_{16} + D_U + \alpha_R M^2_{1/2} 
                               - s^2_W M^2_Z \cos 2 \beta       
                      - 2 \Delta_\tau          + m^2_\tau                    \\
 m^2_{\tilde{\tau}_L} &= m^2_{16} - 3 D_U + \alpha_L M^2_{1/2}
               - (1/2 - s^2_W) M^2_Z \cos 2 \beta
                      - \Delta_\tau - \Delta_{\nu_\tau} + m^2_\tau,        \\
 m^2_{\tilde{\nu}_{\tau L}} &= m^2_{16} - 3 D_U + \alpha_L M^2_{1/2} 
                                      + 1/2 M^2_Z \cos 2 \beta     
                       - \Delta_\tau - \Delta_{\nu_\tau}. 
\label{eq:mtausnu}
\end{align}
[The factor 2 in front of $\Delta_\tau$ in Eq.~(\ref{eq:mtaur}) is due to
the fact that the $SU(2)$ doublet is propagating in the loop.]
The contribution $\Delta_{\nu_\tau}$ carries the information on the value
of the heavy right-handed neutrino mass. As will be shown below, 
this parameter is proportional 
to the Yukawa coupling squared in the $\nu_\tau$ sector which, 
within the seesaw  mechanism,
is linear in the heavy neutrino mass of the third generation. 
Since both parameters $\Delta_{\tau}$ and $\Delta_{\nu_\tau}$
are positive, the masses in the stau sector are shifted downwards compared to
the masses in the selectron sector. The shift is enhanced by the contribution
$\Delta_{\nu_\tau}$ related to the heavy tau neutrino.

In the specific $SO(10)$ model we analyze in the present context, the
two shifts are numerically given by
\begin{equation}
\Delta_\tau = 0.64 \times 10^3 \gev^2 \;\;\; {\rm{and}} \;\;\;
\Delta_{\nu_\tau} = 4.07 \times 10^3 \gev^2.   
\end{equation}
These shifts are individually significantly larger than the typical errors
of the slepton/sneu\-trino masses squared 
which are of order $ 0.1 \;{\rm{to}}\; 0.2 \times 10^3$ GeV$^2$. The effects of
both the tau and the neutrino tau Yukawa coupling can therefore be extracted
experimentally.\\

We shall exploit these relations systematically to determine the GUT
scale parameters and the right-handed neutrino mass in the framework
of the error estimates for the charged and neutral sleptons of the
first and third generation.

\paragraph{(a) Electroweak SM D-terms} \ \\[1mm]
The difference between $e$-sneutrino and L-selectron mass 
\begin{align}
   m^2_{\tilde{\nu}_{e\LL}} -  m^2_{\tilde{e}_\LL} &= 
       D^{elw}_{\tilde{\nu}_\LL} - D^{elw}_{\tilde{e}_\LL}  \\  
	&= \MW^2 \cos 2\beta
\end{align}
measures the familiar D-terms associated with the SU(3)$\times$SU(2)$\times$U(1) 
symmetry breaking in the Standard Model. 
Radiative corrections are incorporated at the one-loop level in the numerical 
evaluation. 
The mass difference predicted by the electroweak D-terms and the radiative corrections,
\begin{equation}
       \MW^2 \cos 2\beta + {\rm{rad.cor.}} = -6.175 \times 10^3 \gev^2,
\end{equation}
is well reproduced by the simulated mass measurements:
\begin{equation}
m^2_{\tilde{\nu}_{e\LL}} -  m^2_{\tilde{e}_\LL} = 
         -6.280^{+.255}_{-.260}\times 10^3 \, \gev^2.
\end{equation}
This relation may serve as a consistency check for measurements. Here and
in the next subsection the sneutrino mass error estimated in section 2 and the 
selectron mass errors 
\begin{equation}
m_{\tilde{e}_\RR} = (125.32 \pm 0.05) \gev \;\;\; {\rm{and}} \;\;\;
m_{\tilde{e}_\LL} = 190.0^{+0.4}_{-0.3} \gev
\end{equation}
from Ref.~\cite{slep} are used.

\paragraph{(b) Universal mass parameter
 \boldmath $m_{16}$ and GUT D-term} \ \\[1mm]
The most precise measurements of the soft mass term $m_{16}$ and the GUT D-term 
$D_U$ can be performed in the charged slepton sector:
\begin{align}
   3 m^2_{\tilde{e}_\RR} + m^2_{\tilde{e}_\LL}  &= 4 m^2_{16} +
         (3 \alpha_R + \alpha_L) M^2_{1/2} - (1/2 + 2 \sw^2) \MZ^2 \cos 2\beta    \\
   m^2_{\tilde{e}_\RR} - m^2_{\tilde{e}_\LL} &= 4 D_U + 
                         (\alpha_R - \alpha_L) M^2_{1/2} + (1/2 - 2 s^2_W)
   	\MZ^2 \cos 2\beta 
\end{align}
The universal mass 
parameter $M_{1/2}$ is assumed to be pre-determined very precisely in the 
gaugino sector of the theory. In the numerical analysis the next-to-leading 
order is included properly. Using the estimated mass measurement errors, 
the fundamental parameters $m_{16}$ and $D_U$ at the high scale can be precisely 
determined:
\begin{equation}
m_{16} = 70.0^{+0.3}_{-0.2} \gev, \qquad
D_U = 0^{+30}_{-40} \gev^2.
\end{equation}
The scalar mass parameter $m_{16}$ can be extracted with an accuracy 
significantly 
better than 1 GeV while the square-root of the SO(10) D-term can be measured 
 at the level of less than 10 GeV.
 
\paragraph{(c) Yukawa interactions and seesaw scale} \ \\[1mm]
As noticed in Refs.~\cite{Baer:2000hx,blair:00p}, the heavy R-neutrino
mass affects the evolution of the iso-doublet scalar mass parameters
above the seesaw scale through the neutrino Yukawa couplings, but the
iso-scalar parameters will be much less affected. Since the effect is
induced by the Yukawa couplings, we anticipate that only the third
generation will signal the seesaw scale.

To simplify the analysis, we eliminate the stau mixing parameters by adding
up the 1 and 2 masses squared, or the L and R masses, correspondingly.
Part of the difference between the stau and the selectron sector
is induced by the {\underline{$\tau$-Yukawa coupling}} as can easily be seen
by inspecting the renormalization group equations (see e.g.~\cite{blair:00p} 
and references therein). This part can be projected out by using the
sum rule
\begin{equation}
2 \Delta_{\tau} = (m^2_{\tilde{e}_\LL} + 
		     m^2_{\tilde{e}_\RR} - 
		     m^2_{\tilde{\nu}_{e \LL}}) -
		  (m^2_{\tilde{\tau}_1} + m^2_{\tilde{\tau}_2} -
		   m^2_{\tilde{\nu}_{\tau \LL}}) + 2 m_\tau^2.
\end{equation}
Inserting the expected experimental errors for the masses from section~\ref{dec}
 and Refs.~\cite{slep,lcws04},
\begin{equation}
 \Delta_{\tau} = 0.63^{+0.28}_{-0.27} \times 10^3 \gev^2.
\end{equation}
We conclude that the effect of the $\tau$ Yukawa coupling 
can be isolated at the 2-$\sigma$ level.\\ 

As we are primarily interested in identifying the 
{\underline{effect of the right-handed neutrinos} 
we consider the following sum rule for the $\Delta_{\nu_\tau}$ parameter
derived from eqs.~(\ref{eq:mselr})-(\ref{eq:mtausnu}):
\begin{equation}
2 \Delta_{\nu_\tau}[M_{\RR_3}] = (3 m^2_{\tilde{\nu}_{e \LL}}
		- m^2_{\tilde{e}_\LL} - m^2_{\tilde{e}_\RR}) -
		  (3 m^2_{\tilde{\nu}_{\tau \LL}} - 
		  m^2_{\tilde{\tau}_1} - m^2_{\tilde{\tau}_2}) - 2 m_\tau^2.
\label{eq:deltaM3}
\end{equation}
This relation holds exactly at tree-level and gets modified 
by small corrections
at the one-loop level. 
The particular form of eq.~(\ref{eq:deltaM3}) 
implies that all D-terms (electroweak and GUT-induced) and the
effects of the $\tau$ Yukawa coupling cancel.
With the simulated mass measurement results one obtains
\begin{equation}
 \Delta_{\nu_\tau} = 4.1^{+0.55}_{-0.56} \times 10^3 \gev^2.
 \label{eq:Deltanutau}
\end{equation}
It follows from the renormalization group equations 
that $\Delta_{\nu_\tau}[M_{\RR_3}]$ is of the order
$ Y^2_\nu \log{M^2_{GUT}/M^2_{\RR_3}}$.
Since the Yukawa coupling $Y_\nu$ can be 
estimated in the seesaw mechanism by the mass 
$m_{\nu_3}$ of the third left-handed neutrino,
\begin{equation}
    Y_\nu^2 = m_{\nu_3} M_{\RR_3} / (v \, \sin\beta)^2,
\end{equation}
the parameter $\Delta_{\nu_\tau}[M_{R_3}]$ depends approximately linearly on
the mass $M_{\RR_3}$,
\begin{equation}
\Delta_{\nu_\tau}[M_{R_3}] \simeq  \frac{m_{\nu_3} M_{\RR_3}}
                                         {16 \pi^2 (v \, \sin\beta)^2}
\left( 3 m^2_{16}  + A^2_0 \right)
\log\frac{M^2_{GUT}}{M^2_{\RR_3}},
\label{eq:Dnuev}
\end{equation}
so that it can well be determined. [$v$ and $\tan\beta$ are the usual
parameters in the Higgs sector.]
Inserting the pre-determined value for $m_{16}$ from the analyses
above and the trilinear coupling $A_0$ from stop mixing \cite{Boos:2003vf}, 
we can calculate $M_{\RR_3}$, cf.~Fig.~\ref{fig:MR}. 
\begin{figure}[tb]
\centering
\epsfig{file=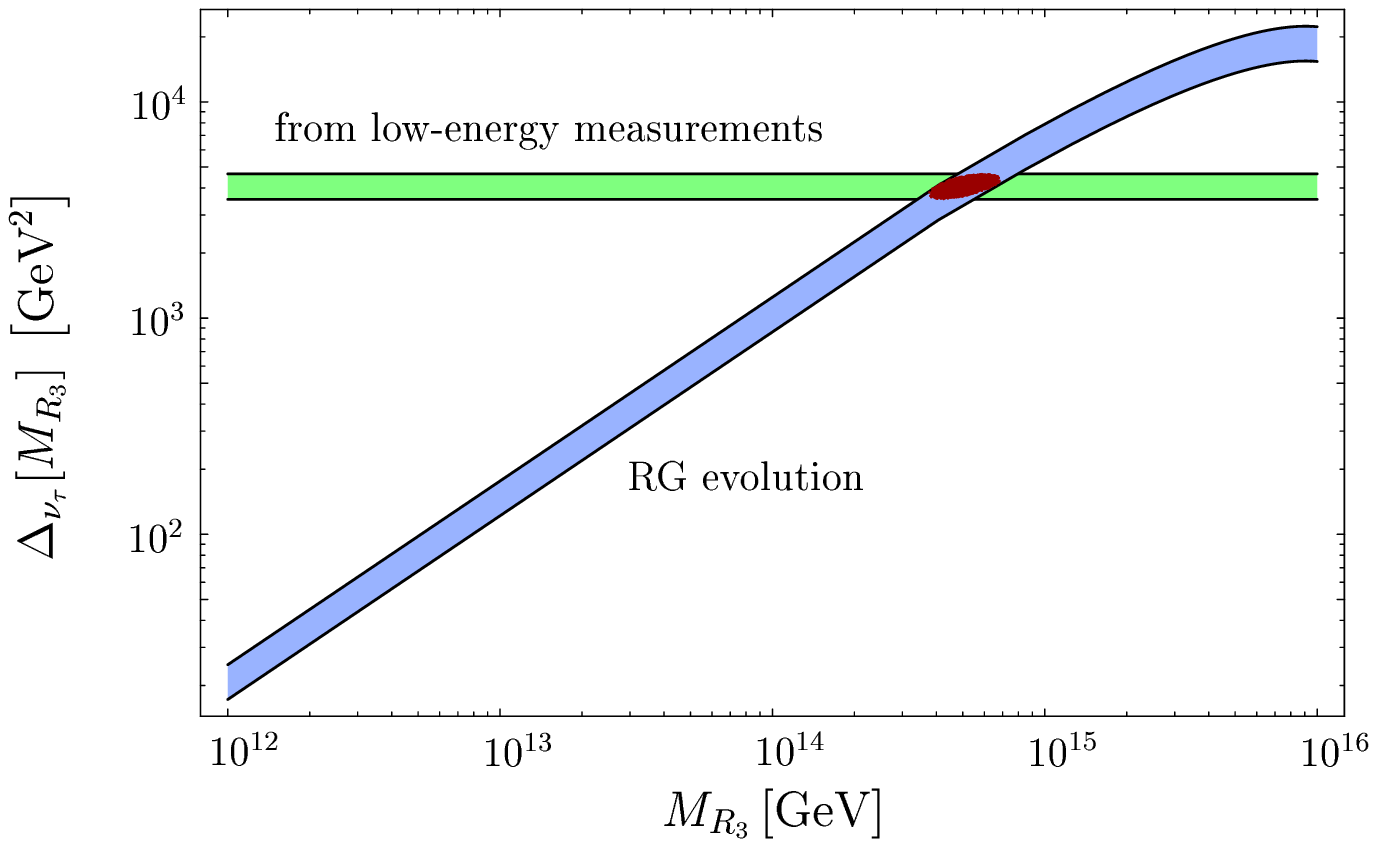,width=5.2in, bb=45 441 435 681, clip=true}
\mycaption{Shift $\Delta_{\nu_\tau}$ 
in the evolution of the tau-neutrino mass as calculated 
from the renormalization group equations,
eq.~\eqref{eq:Dnuev} (blue band) and compared with
low-energy mass measurements, eq.~\eqref{eq:Deltanutau} (green band).
The widths of the bands indicate estimated
one-standard-deviations errors of the experimental input parameters. The red
crossing region is the statistical combination which determines the 
neutrino seesaw scale
$M_{\RR_3}$ of the third generation.}
\label{fig:MR}
\end{figure}
Assuming hierarchical neutrino masses,
we have identified the error of the
neutrino mass $m_{\nu_3}$ with the current
uncertainty of the atmospheric neutrino mass difference \cite{nufit}. 
With these assumptions, one obtains
\begin{equation}
  M_{\RR_3} = 3.7 \; {\rm{to}} \; 6.9 \times 10^{14} \gev,
\end{equation}
to be compared with the initial value $M_{\RR_3} = 6 \times 10^{14}$ GeV. 
This analysis thus provides us with a unique estimate of the 
high-scale $\nu_R$ mass parameter $M_{\RR_3}$.  


\paragraph{Added Note} \ \\[2mm]
As mentioned in footnote 4,  we have up to now neglected
the effect of lepton flavor violation (LFV)
induced in the soft SUSY breaking parameters of the slepton sector
due to the effect of neutrino Yukawa interactions in the RGE evolution
of the parameters. These in turn lead to flavor violating lepton decays
such as $\mu \to e \gamma$ \cite{Borzumati:1986qx}.
However, it has been shown in Ref.~\cite{Masiero:2002jn} that for the
model under consideration the bounds due to rare lepton decays
are not violated: BR$(\mu \to e \gamma) \sim O(10^{-13})$ and
BR$(\tau \to \mu \gamma) \sim O(10^{-8})$.
The latter is close to the latest experimental bound of $6.8 \times 10^{-8}$
\cite{Aubert:2005ye} and it will be tested at the $B$ factories within the next
few years.
The former branching ratio will be scrutinized in a dedicated PSI experiment.
A second class of observables
can be determined in
flavor violating decays of supersymmetric particles,
see \cite{Deppisch:2003wt} and further references quoted in these papers. 
In particular, the off-diagonal terms in $M^2_L$ imply that every sneutrino 
can decay in principle into all charged leptons.
However, when assuming that
$Y_\nu$ $(Y_l)$ has a structure similar to $Y_u$ $(Y_d)$ at $M_{GUT}$
as in the model under consideration,
the flavor violation decays are suppressed by powers of
the CKM-matrix $V_{ij}$. This implies that our results are only weakly affected
by the additional decay modes and the LFV decays reach at most 10\%.
For the same reason the production
of $\tilde \nu_\mu$ and $\tilde \nu_\tau$ in $e^- \gamma$ reactions
is expected to be small.
Clearly, the measurements of LFV sneutrino and slepton decays will
provide additional information on the effects generated by $Y_\nu$ and,
thus, can be used to scrutinize the physics at the see-scale $M_{\RR_3}$
further.


\section{Summary}
In this report we have shown that chargino decays to sneutrinos and
charged leptons, if kinematically allowed as two-body decays, provide
an excellent opportunity to measure the masses of the L-sneutrinos
in the three generations. Accuracies better than a per-cent can be
expected when these masses are measured in chargino pair production
at an $e^+e^-$ linear collider. [R-sneutrinos acquire masses near
the GUT scale by the seesaw mechanism.] Independent cross-checks
may be performed in the production of $e$-sneutrinos and charginos
in high energy $e\gamma$ collisions.

Of particular interest is the comparison of scalar masses in the tau
and the electron sector. If the scalar mass parameters are universal
at the GUT scale, as in minimal supergravity, this regularity can be
unraveled in the first and second generation of the scalar masses at the
electroweak scale. However, universality will be broken between the
first two and the third generation in theories incorporating the seesaw
mechanism. The running of the masses from the GUT to the electroweak
scale will be affected by loops involving the heavy R-neutrino with
large Yukawa coupling in the third generation. Sum rules for mass
differences of sneutrinos and selectrons between the first and third
generation can be constructed that project out this contribution.
Being approximately linear in the seesaw scale, the scale can be
estimated from the sneutrino and slepton masses with good accuracy.
In this way a method has been found by which the high R-neutrino mass
can, at least indirectly, be measured. Thus, by means
of extrapolations governed by the renormalization group,
the high accuracy in the
slepton and sneutrino mass measurements can be exploited to determine
high-scale parameters that cannot be accessed directly.


\section*{Acknowledgments}

We are grateful to H.-U.~Martyn and D.~E.~Zerwas for numerous 
discussions on experimental aspects and background estimates.
Moreover, special thanks go to H.-U.~Martyn for
his critical reading of the manuscript.
W.P.~is supported by a MCyT Ramon y Cajal contract,
by the Spanish grant BFM2002-00345, by the
European Commission Human Potential Program RTN network
HPRN-CT-2000-00148  and 
partly by the Swiss 'Nationalfonds'. P.M.Z. thanks 
M.E.~Peskin of SLAC and DFG for partial support; he gratefully 
acknowledges the warm hospitality extended to him during 
a stay at the Stanford Linear Accelerator Center SLAC 
where part of this study was worked out.



\vspace*{3mm}
\begin{thebibliography}{99}
\frenchspacing

\bibitem{Babu:1992ia}
  K.~S.~Babu and R.~N.~Mohapatra,
  Phys.\ Rev.\ Lett.\  {\bf 70} (1993) 2845;\\
for a review see: 
  S.~F.~King,
  Rept.\ Prog.\ Phys.\  {\bf 67} (2004) 107.

\bibitem{seesaw} 
  P.~Minkowski,
  Phys.\ Lett.\ B {\bf 67} (1977) 421;\\
%
M.~Gell-Mann, P.~Ramond, and R.~Slansky, 
in {\it Proc. of the Workshop on Complex Spinors and Unified Theories}, 
Stony Brook, New York, North-Holland, 1979;\\
%
T.~Yanagida, (KEK, Tsukuba), 1979;\\
%
R.~N.~Mohapatra and G.~Senjanovic,
Phys.\ Rev.\ Lett.\  {\bf 44} (1980) 912.


\bibitem{sneu}
  A.~Freitas, A.~von Manteuffel and P.~M.~Zerwas,
  Eur.\ Phys.\ J.\ C {\bf 40}, 435 (2005).

\bibitem{nauenberg}
U.~Nauenberg,
Contribution to the {\it 3rd Workshop of the Extended ECFA/DESY Linear Collider
Study}, Prague, Czech Republic
(2002).

\bibitem{bhk}
  V.~D.~Barger, T.~Han and J.~Kelly,
  Phys.\ Lett.\ B {\bf 419} (1998) 233.

\bibitem{sps}
B.~C.~Allanach {\it et al.},
Eur.\ Phys.\ J.\ C {\bf 25} (2002) 113.

\bibitem{spa}
Supersymmetry Parameter Analysis (SPA) Project,
\texttt{http://spa.desy.de/spa/}.

\bibitem{lcreports} 
J.~A.~Aguilar-Saavedra {\it et al.}  [ECFA/DESY LC Physics Working Group],
arXiv:hep-ph/0106315; \\
%
T.~Abe {\it et al.}  [American Linear Collider Working Group],
in {\it Proc. of the APS/DPF/DPB Summer Study on the Future of Particle Physics (Snowmass 2001) } ed. N.~Graf,
  arXiv:hep-ex/0106056; \\
  K.~Abe {\it et al.}  [ACFA Linear Collider Working Group],
  arXiv:hep-ph/0109166.

\bibitem{choi:00}
S.~Y.~Choi, A.~Djouadi, M.~Guchait, J.~Kalinowski, H.~S.~Song and P.~M.~Zerwas,
Eur.\ Phys.\ J.\ C {\bf 14} (2000) 535.

\bibitem{slep}
A.~Freitas, A.~von Manteuffel and P.~M.~Zerwas,
Eur.\ Phys.\ J.\ C {\bf 34} (2004) 487.

\bibitem{Zstop}
J.~H.~K\"uhn, A.~Reiter and P.~M.~Zerwas,
Nucl.\ Phys.\ B {\bf 272} (1986) 560.

\bibitem{CMZ}
S.~Y.~Choi, H.-U.~Martyn and P.~M.~Zerwas, DESY 05-150 and hep-ph/0508021 [Eur.\ Phys.\ J.\
{\bf C} {\it in press}].

\bibitem{lhclc}
LHC/LC Study Group Working Report, eds. G.~Weiglein {\it et al.}, 
hep-ph/0410364;\\
K.~Desch, J.~Kalinowski, G.~Moortgat-Pick, M.~M.~Nojiri and G.~Polesello,
JHEP {\bf 0402}, 035 (2004).
 
\bibitem{martyn}
H.-U.~Martyn, LC Note LC-PHSM-2003-071,
  hep-ph/0406123.
 
\bibitem{ginz}
  I.~F.~Ginzburg, G.~L.~Kotkin, V.~G.~Serbo and V.~I.~Telnov,
  JETP Lett.\  {\bf 34} (1981) 491
  [Pisma Zh.\ Eksp.\ Teor.\ Fiz.\  {\bf 34} (1981) 514].

\bibitem{telnov}
V.~A.~Telnov, {\it A Code for the Simulation of Luminosities and QED Backgrounds
at Photon Colliders,} presented at the ECFA-DESY Linear Collider Workshop, St.
Malo, France, April 2002.

\bibitem{compaz}
  A.~F.~Zarnecki,
  Acta Phys.\ Polon.\ B {\bf 34} (2003) 2741.

\bibitem{Kolda:1995iw}
C.~F.~Kolda and S.~P.~Martin,
Phys.\ Rev.\ D {\bf 53} (1996) 3871.
 
\bibitem{Buchmuller:2001dc}
W.~Buchm\"uller and D.~Wyler,
Phys.\ Lett.\ B {\bf 521} (2001) 291.

\bibitem{Masiero:2002jn}
A.~Masiero, S.~K.~Vempati and O.~Vives,
Nucl.\ Phys.\ B {\bf 649} (2003) 189.

\bibitem{blair:00p}
G.~A.~Blair, W.~Porod and P.~M.~Zerwas,
Eur.\ Phys.\ J.\ C {\bf 27} (2003) 263.

\bibitem{lcws04}
A.~Freitas, H.~U.~Martyn, U.~Nauenberg and P.~M.~Zerwas,
  in {\it Proc. of the International Conference on Linear Colliders (LCWS 04), 
  Paris, France, 19-24 Apr 2004}
  [hep-ph/0409129].

\bibitem{Baer:2000hx}
H.~Baer, C.~Balazs, J.~K.~Mizukoshi and X.~Tata,
Phys.\ Rev.\ D {\bf 63} (2001) 055011.

\bibitem{Boos:2003vf}
  E.~Boos, H.~U.~Martyn, G.~Moortgat-Pick, M.~Sachwitz, A.~Sherstnev and P.~M.~Zerwas,
  Eur.\ Phys.\ J.\ C {\bf 30} (2003) 395; \\
 A.~Bartl, H.~Eberl, S.~Kraml, W.~Majerotto, W.~Porod and A.~Sopczak,
  Z.\ Phys.\ C {\bf 76} (1997) 549.

\bibitem{nufit}
M.~Maltoni, T.~Schwetz, M.~A.~Tortola and J.~W.~F.~Valle,
New J.\ Phys.\  {\bf 6} (2004) 122; \\
%
  G.~L.~Fogli, E.~Lisi, A.~Marrone and A.~Palazzo,
  hep-ph/0506083.

\bibitem{Borzumati:1986qx}
  F.~Borzumati and A.~Masiero,
  Phys.\ Rev.\ Lett.\  {\bf 57} (1986) 961;\\
  J.~Hisano, T.~Moroi, K.~Tobe and M.~Yamaguchi,
  Phys.\ Rev.\ D {\bf 53} (1996) 2442;\\
 F.~Deppisch, H.~P\"as, A.~Redelbach, R.~R\"uckl and Y.~Shimizu,
  Eur.\ Phys.\ J.\ C {\bf 28} (2003) 365.

\bibitem{Aubert:2005ye}
B.~Aubert {\it et al.}  [BABAR Collaboration],
hep-ex/0502032.

\bibitem{Deppisch:2003wt}
N.~Arkani-Hamed, H.~C.~Cheng, J.~L.~Feng and L.~J.~Hall,
Phys.\ Rev.\ Lett.\  {\bf 77}, 1937 (1996) and
Nucl.\ Phys.\ B {\bf 505}, 3 (1997);\\
  J.~Hisano, M.~M.~Nojiri, Y.~Shimizu and M.~Tanaka,
Phys.\ Rev.\ D {\bf 60}, 055008 (1999);
  W.~Porod and W.~Majerotto,
  Phys.\ Rev.\ D {\bf 66} (2002) 015003;\\
F.~Deppisch, H.~P\"as, A.~Redelbach, R.~R\"uckl and Y.~Shimizu,
Phys.\ Rev.\ D {\bf 69} (2004) 054014;\\
F.~Deppisch, H.-U.~Martyn, H.~P\"as, A.~Redelbach and R.~R\"uckl,
  in {\it Proc. of the International Conference on Linear Colliders (LCWS 04), 
  Paris, France, 19-24 Apr 2004}
[hep-ph/0408140];\\
W.~Porod,
hep-ph/0410318.

\end{thebibliography}
\end{document}